# Improper statistics of the radiation from a randomly rotating source.

**MIKHAIL CHARNOTSKII**

*Erie. CO, 80516, USA*
*\*Mikhail.Charnotskii@gmail.com*

**Abstract:** Our earlier work [J. Opt. **19**. (2017) 0905603] showed that, in contrast to the four parameters of the traditional Stokes vector description of the statistics of the partially polarized light, the complete second-order statistics of the narrow-band polarized wave is characterized by thirteen parameters. Here we analyze the second-order statistics of the radiation from a randomly rotating source of the electromagnetic radiation, and show that it includes a covariance of the right and left circular polarizations that is not captured by the Stokes vector formalism. We illustrate this finding on a simple example of rotating quadrupole.

1. **Introduction**

Traditionally the Stokes vector [1] has been exclusively used to describe the local polarization state of electromagnetic radiation. The popularity of the Stokes parameters is because a linear polarizer and retardation plate are all that is needed to measure them [2]. The more recent development of the unified theory of coherence and polarization of stochastic electromagnetic beams, see for example [3], added spatial coherence to the polarization theory and made it possible to investigate the propagation of partially polarized beam waves. However, at each point the covariance matrix of the field, which is also known as a cross-spectral density matrix contains the same information as the Stokes vector.

In the recent paper [5] we examined a complete second-order statistics of the narrow-band partially polarized light and showed that, in general, it is characterized by thirteen independent real parameters, whereas the classic Stokes vector description utilizes only four parameters. General statistics developed in [5] allows for the non-circular random complex amplitudes, hence improper signals, [6]. Meanwhile the traditional Stokes or covariance matrix description implies circular complex amplitudes and proper signals. However, a coherent receiver is needed to measure some of these additional, non-Stokes, parameters. The objective of this paper is to present practical examples of partially polarized field that have non-zero Exo-Stokes parameters, and demonstrate that they carry additional information about the source, which is hidden from the Stokes description.

The paper is organized as follows. In section 2 we review the main results of [5] regarding the complete second-order statistics of the partially polarized field. In section 3 we develop the second order statistics for the far-field field created by a randomly rotated coherent source of polarized radiation. In section 4 we illustrate the general results of the previous section on simple examples of the dipole and quadrupole sources. Section 5 summarizes the findings.

2. **Second-order statistics of the partially polarized field**

Electric field $\mathbf{E}(t)$ of the transverse electromagnetic wave with carrier frequency $\omega$ $\omega$ with local wave front normal along the *z*-axis at some point in the $(x, y)$ plane can be presented as

$$\mathbf{E}(t) = E_X(t)\hat{\mathbf{x}} + E_Y(t)\hat{\mathbf{y}} = \text{Re}\left[u(t)e^{i\omega t}\hat{\mathbf{x}}\right] + \text{Re}\left[v(t)e^{i\omega t}\hat{\mathbf{y}}\right]. \quad \textbf{(1)}$$

Here $E_X$ and $E_Y$ are orthogonal components of an electrical field in the *x* and *y* directions, $u(t)$ and $v(t)$ are random complex amplitudes of these components. For the quasi-monochromatic waves considered here, correlation time $t_C$ of the $u(t)$ and $v(t)$ is much larger than the carrier oscillation period

$$\omega t_C \gg 1. \tag{2}$$

Field $\mathbf{E}(t)$ is a real-valued two-dimensional nonstationary random vector that can be represented as

$$\mathbf{E}(t) = \left[ u_R(t)\cos(\omega t) - u_I(t)\sin(\omega t) \right]\hat{\mathbf{x}} + \left[ v_R(t)\cos(\omega t) - v_I(t)\sin(\omega t) \right]\hat{\mathbf{y}}, \tag{3}$$

where subscripts $R$ and $I$ stand for the real and imaginary parts (in-phase and quadrature components) of corresponding complex amplitudes.

Presenting the in-phase and quadrature parts of the complex amplitudes as sums of mean and the fluctuating components

$$\begin{aligned} u_R(t) &= \langle u_R \rangle + \tilde{u}_R(t), \; u_I(t) = \langle u_I \rangle + \tilde{u}_I(t), \\ v_R(t) &= \langle v_R \rangle + \tilde{v}_R(t), \; v_I(t) = \langle v_I \rangle + \tilde{v}_I(t), \end{aligned} \tag{4}$$

we limit our attention to the mean values $\langle u_R \rangle, \langle u_I \rangle, \langle v_R \rangle$ and $\langle v_I \rangle$, and the covariance matrix of the fluctuating components of the Jones vector

$$\mathbf{C} = \begin{pmatrix} \langle \tilde{u}_R^2 \rangle & \langle \tilde{u}_R \tilde{u}_I \rangle & \langle \tilde{u}_R \tilde{v}_R \rangle & \langle \tilde{u}_R \tilde{v}_I \rangle \\ \langle \tilde{u}_R \tilde{u}_I \rangle & \langle \tilde{u}_I^2 \rangle & \langle \tilde{u}_I \tilde{v}_R \rangle & \langle \tilde{u}_I \tilde{v}_I \rangle \\ \langle \tilde{u}_R \tilde{v}_R \rangle & \langle \tilde{u}_I \tilde{v}_R \rangle & \langle \tilde{v}_R^2 \rangle & \langle \tilde{v}_R \tilde{v}_I \rangle \\ \langle \tilde{u}_R \tilde{v}_I \rangle & \langle \tilde{u}_I \tilde{v}_R \rangle & \langle \tilde{v}_R \tilde{v}_I \rangle & \langle \tilde{v}_I^2 \rangle \end{pmatrix}. \tag{5}$$

Here we do not require the complex amplitudes $u$ and $v$ to be circular or proper, [6]. Mean values $\langle u_R \rangle, \langle u_I \rangle, \langle v_R \rangle, \langle v_I \rangle$ describe the coherent elliptically polarized component of the field, [5]. Covariance matrix $\mathbf{C}$ fully describes the single point second-order statistics of the field fluctuations. Any other second-order field statistic can be represented in these terms. The source model considered further on leads to the zero-mean field, and we exclude it from the further formulations for brevity. Several examples of the partially polarized fields including coherent components have been presented in [5].

Stokes vector is represented in terms of the elements of the covariance matrix $\mathbf{C}$, Eq. (5) as

$$\mathbf{S} = \begin{pmatrix} \langle u^2 \rangle + \langle v^2 \rangle \\ \langle u^2 \rangle - \langle v^2 \rangle \\ 2\,\mathrm{Re}\langle uv^* \rangle \\ -2\,\mathrm{Im}\langle uv^* \rangle \end{pmatrix} = \begin{pmatrix} \langle \tilde{u}_R^2 \rangle + \langle \tilde{u}_I^2 \rangle + \langle \tilde{v}_R^2 \rangle + \langle \tilde{v}_I^2 \rangle \\ \langle \tilde{u}_R^2 \rangle + \langle \tilde{u}_I^2 \rangle - \langle \tilde{v}_R^2 \rangle - \langle \tilde{v}_I^2 \rangle \\ 2\langle \tilde{u}_R \tilde{v}_R \rangle + 2\langle \tilde{u}_I \tilde{v}_I \rangle \\ 2\langle \tilde{u}_R \tilde{v}_I \rangle - 2\langle \tilde{u}_I \tilde{v}_R \rangle \end{pmatrix}, \tag{6}$$

and the covariance matrix (mutual intensity) of Jones vector is

$$\mathbf{W} = \begin{pmatrix} \langle uu^* \rangle & \langle uv^* \rangle \\ \langle u^*v \rangle & \langle vv^* \rangle \end{pmatrix} = \begin{pmatrix} \langle \tilde{u}_R^2 \rangle + \langle \tilde{u}_I^2 \rangle & \begin{array}{c} \langle \tilde{u}_R \tilde{v}_R \rangle + \langle \tilde{u}_I \tilde{v}_I \rangle \\ -i\langle \tilde{u}_R \tilde{v}_I \rangle + i\langle \tilde{u}_I \tilde{v}_R \rangle \end{array} \\ \begin{array}{c} \langle \tilde{u}_R \tilde{v}_R \rangle + \langle \tilde{u}_I \tilde{v}_I \rangle \\ +i\langle \tilde{u}_R \tilde{v}_I \rangle - i\langle \tilde{u}_I \tilde{v}_R \rangle \end{array} & \langle \tilde{v}_R^2 \rangle + \langle \tilde{v}_I^2 \rangle \end{pmatrix}. \tag{7}$$

Both Stokes vector and covariance matrix **W** have only four independent parameters leaving the extra six parameters of the covariance matrix **C** unaccounted for. These six parameters are captured by so-the called complementary-covariance matrix, [6].

$$\mathbf{W}_C = \begin{pmatrix} \langle u^2 \rangle & \langle uv \rangle \\ \langle uv \rangle & \langle v^2 \rangle \end{pmatrix} = \begin{pmatrix} \langle \tilde{u}_R^2 \rangle - \langle \tilde{u}_I^2 \rangle + 2i\langle \tilde{u}_I \tilde{u}_I \rangle & \langle \tilde{u}_R \tilde{v}_R \rangle - \langle \tilde{u}_I \tilde{v}_I \rangle \\ & +i\langle \tilde{u}_R \tilde{v}_I \rangle + i\langle \tilde{u}_I \tilde{v}_R \rangle \\ \langle \tilde{u}_R \tilde{v}_R \rangle - \langle \tilde{u}_I \tilde{v}_I \rangle & \langle \tilde{v}_R^2 \rangle - \langle \tilde{v}_I^2 \rangle + 2i\langle \tilde{v}_R \tilde{v}_I \rangle \\ +i\langle \tilde{u}_R \tilde{v}_I \rangle + i\langle \tilde{u}_I \tilde{v}_R \rangle & \end{pmatrix}. \quad (8)$$

Complementary-covariance matrix $\mathbf{W}_C$ carries the remaining six parameters of the correlation matrix **C**. Proper complex random vectors have identically zero pseudo-covariance matrix [6], and Stokes vector or covariance matrix are sufficient for their second-order statistics, provided that coherent field is absent. Proper complex random vectors are essentially the wide-sense weakly circular random vectors, [7]. Most of the polarization studies, sometimes implicitly, make the circularity assumption and limits its scope to the covariance matrix **W** or Stokes vector, [8]. The complete second-order statistics of the partially polarized light including complementary covariance matrix was developed in [5]. Here we present a simple physical example of a naturally emerging improper random polarized field.

### 3. Statistics of the field from randomly rotating source

We present the transverse electrical field from a coherent source in the radiation zone as

$$\mathbf{E}(\mathbf{r}) = \operatorname{Re}\left[\frac{1}{r}\mathbf{S}(\alpha,\beta)\exp(ikr - i\omega t)\right]. \quad (8)$$

Here we use the source coordinate system in Cartesian $(x,y,z)$ or spherical $(r,\alpha,\beta)$ formulation where the position vector **r** of the observation point is

$$\mathbf{r} = x\hat{\mathbf{x}} + y\hat{\mathbf{y}} + z\hat{\mathbf{z}} = r(\cos\alpha\cos\beta\hat{\mathbf{x}} + \sin\alpha\cos\beta\hat{\mathbf{y}} + \sin\beta\hat{\mathbf{z}}) = r\hat{\mathbf{r}}. \quad (9)$$

It is convenient to use the basis vectors of the spherical coordinate system

$$\begin{aligned}\hat{\mathbf{r}} &= \cos\alpha\cos\beta\hat{\mathbf{x}} + \sin\alpha\cos\beta\hat{\mathbf{y}} + \sin\beta\hat{\mathbf{z}}, \\ \hat{\boldsymbol{\alpha}} &= -\sin\alpha\hat{\mathbf{x}} + \cos\alpha\hat{\mathbf{y}}, \\ \hat{\boldsymbol{\beta}} &= \cos\alpha\sin\beta\hat{\mathbf{x}} + \sin\alpha\sin\beta\hat{\mathbf{y}} - \cos\beta\hat{\mathbf{z}}\end{aligned} \quad (10)$$

to warrant the transverse property of the field by presenting the vector radiation pattern $\mathbf{S}(\alpha,\beta)$ as

$$\mathbf{S}(\alpha,\beta) = S_\alpha(\alpha,\beta)\hat{\boldsymbol{\alpha}} + S_\beta(\alpha,\beta)\hat{\boldsymbol{\beta}} \quad (11)$$

We use the Euler rotation matrix

$$\mathbf{R}(\alpha,\beta,\gamma) = \begin{pmatrix} \cos\alpha\sin\beta\cos\gamma - \sin\alpha\sin\gamma & \sin\alpha\sin\beta\cos\gamma + \cos\alpha\sin\gamma & -\cos\beta\cos\gamma \\ -\cos\alpha\sin\beta\sin\gamma - \sin\alpha\cos\gamma & -\sin\alpha\sin\beta\sin\gamma + \cos\alpha\cos\gamma & \cos\beta\sin\gamma \\ \cos\alpha\cos\beta & \sin\alpha\cos\beta & \sin\beta \end{pmatrix} \quad (12)$$

to bring the observation point to the fixed receiver position at $\mathbf{r} = r\hat{\mathbf{z}}$. Here azimuthal angle $\alpha$ and elevation angle $\beta$ determine the direction of the nominal source axis, and angle $\gamma$ describes rotation around this axis. After the same rotation the field vector $\mathbf{S}(\alpha,\beta)$ in the receiver coordinates is

$$\mathbf{S}(\alpha,\beta) = u(\alpha,\beta,\gamma)\hat{\mathbf{x}} + v(\alpha,\beta,\gamma)\hat{\mathbf{y}}, \qquad (13)$$

where components of the Jones vector in the receiver coordinate system are

$$u(\alpha,\beta,\gamma) = \left[S_\alpha(\alpha,\beta)\cos\gamma + S_\beta(\alpha,\beta)\sin\gamma\right]\frac{e^{ikr}}{r},$$
$$v(\alpha,\beta,\gamma) = \left[-S_\alpha(\alpha,\beta)\sin\gamma + S_\beta(\alpha,\beta)\cos\gamma\right]\frac{e^{ikr}}{r}. \qquad (14)$$

These equations have a clear physical meaning. Receiver "sees" the source from the direction determined by the polar $\alpha$ and elevation $\beta$ angles, and the third Euler angle $\gamma$ describes the source rotation about the observation direction.

Now we need to make some assumptions regarding the detection process. As described in [5], the incident field is split in two channels and passed through linear polarizers to separate the $u$ and $v$ components. These components are mixed with the local oscillator before they are detected. The detector photo current at the intermediate (radio) frequency is down converted and lowpass filtered to recover the complex envelopes $u(t)$ and $v(t)$.

In order to estimate the statistics, these baseband signals must be recorded over detection time $t_D \gg t_C$. In case of the moving source, we assume that relative change of the distance $r$ between the source and receiver is negligible during this time. We also assume that the radial velocity of the source remains constant during the measurement time. This causes the Doppler shift which is handled at the down conversion step by adjusting the oscillator frequency. Still the $ikr$ phase and the phase of the local oscillator remain unknown, and both complex amplitudes $u(t)$ and $v(t)$ are known up to a constant phase term $\exp(i\theta)$.

Under these assumptions we include the $1/r$ factor in the definition of the complex amplitude of the source and present the Jones vector components measured by coherent receiver as

$$u(\alpha,\beta,\gamma) = \left[s_\alpha(\alpha,\beta)\cos\gamma + s_\beta(\alpha,\beta)\sin\gamma\right]e^{i\theta},$$
$$v(\alpha,\beta,\gamma) = \left[-s_\alpha(\alpha,\beta)\sin\gamma + s_\beta(\alpha,\beta)\cos\gamma\right]e^{i\theta}. \qquad (15)$$

We calculate the second-order statistics of the Jones vector for the simple case of completely random source rotation when the Euler angles are statistically independent. Direction of the source axis, determined by the angles $\alpha$ and $\beta$ is uniformly distributed over the unit sphere, and rotation angle $\gamma$ is $[-\pi, \pi]$ uniformly distributed. Thus, probability density function of the random triplet $(\alpha, \beta, \gamma)$ is

$$P(\alpha,\beta,\gamma) = \frac{\cos\beta}{8\pi^2}, \quad -\pi < \alpha \leq \pi,\ -\frac{\pi}{2} \leq \alpha \leq \frac{\pi}{2},\ -\pi < \gamma \leq \pi. \qquad (16)$$

It is clear from Eq. (14) and Eq. (15) that the mean field is zero. This is to be expected, since completely random source rotation cannot allow for any preferred direction or circulation seen by the receiver.

It is more convenient to perform the second moments calculations using the right/left polarization basis rather than the Cartesian basis used up to this point. We introduce the right/left complex unit vectors as

$$\hat{\boldsymbol{\rho}} = \frac{1}{\sqrt{2}}(\hat{\mathbf{x}} + i\hat{\mathbf{y}}),\ \hat{\boldsymbol{\lambda}} = \frac{1}{\sqrt{2}}(\hat{\mathbf{x}} - i\hat{\mathbf{y}}). \qquad (17)$$

The radiated field, Eq. (11) is now presented as

$$\mathbf{S}(\alpha,\beta) = S_R(\alpha,\beta)\hat{\mathbf{\rho}} + S_L(\alpha,\beta)\hat{\mathbf{\lambda}}, \qquad (18)$$

where

$$S_\rho(\alpha,\beta) = \frac{1}{\sqrt{2}}\left[S_\alpha(\alpha,\beta) + iS_\beta(\alpha,\beta)\right], \quad S_\lambda(\alpha,\beta) = \frac{1}{\sqrt{2}}\left[S_\alpha(\alpha,\beta) - iS_\beta(\alpha,\beta)\right] \qquad (19)$$

are the complex amplitudes of the right and left circular polarized components of the radiated field.

Instantaneous field incident at the receiver, Eq. (13) in terms of the right and left circular polarizations is presented as

$$\mathbf{S}(\alpha,\beta) = R(\alpha,\beta,\gamma)\hat{\mathbf{\rho}} + L(\alpha,\beta,\gamma)\hat{\mathbf{\lambda}}, \qquad (20)$$

where complex amplitudes of the circular components are

$$R(\alpha,\beta,\gamma) = \frac{1}{\sqrt{2}}\left[S_\alpha(\alpha,\beta)e^{-i\gamma} + S_\beta(\alpha,\beta)e^{i\gamma}\right]\frac{e^{ikr}}{r} = S_R(\alpha,\beta)e^{-i\gamma}\frac{e^{ikr}}{r},$$

$$L(\alpha,\beta,\gamma) = \frac{1}{\sqrt{2}}\left[S_\alpha(\alpha,\beta)e^{i\gamma} + S_\beta(\alpha,\beta)e^{-i\gamma}\right]\frac{e^{ikr}}{r} = S_L(\alpha,\beta)e^{i\gamma}\frac{e^{ikr}}{r}. \qquad (21)$$

Now we consider a two channels coherent optical receiver that uses circular polarizers instead of the linear polarizers. Similar to Eq. (15), complex amplitudes of the circular components at the output of this receiver can be presented as

$$R(\alpha,\beta,\gamma) = s_R(\alpha,\beta)\exp(i\theta - i\gamma),$$
$$L(\alpha,\beta,\gamma) = s_L(\alpha,\beta)\exp(i\theta + i\gamma). \qquad (22)$$

Eq. (22) is the equivalent of the Eq. (15), but it is more convenient for the further calculations. Mean value of these complex amplitudes are zero, since

$$\langle R \rangle = \int_{-\pi}^{\pi} d\alpha \int_{-\pi/2}^{\pi/2} d\beta \int_{-\pi}^{\pi} d\gamma P(\alpha,\beta,\gamma) R(\alpha,\beta,\gamma)$$
$$= \frac{1}{8\pi^2} \int_{-\pi}^{\pi} d\alpha \int_{-\pi/2}^{\pi/2} d\beta \cos\beta\, s_R(\alpha,\beta) \int_{-\pi}^{\pi} d\gamma \exp(i\theta - i\gamma) = 0, \qquad (23)$$

and similar to the second component.

Covariance matrix of the circular amplitudes is readily calculated as

$$\mathbf{U} = \left\langle \begin{pmatrix} R \\ L \end{pmatrix} (R^* \; L^*) \right\rangle = \begin{pmatrix} \overline{R^2} & 0 \\ 0 & \overline{L^2} \end{pmatrix}, \qquad (24)$$

and complementary covariance matrix, analogous to Eq. (5) is

$$\mathbf{U}_C = \left\langle \begin{pmatrix} R \\ L \end{pmatrix} (R \; L) \right\rangle = \begin{pmatrix} 0 & \overline{RL}e^{2i\theta} \\ \overline{RL}e^{2i\theta} & 0 \end{pmatrix}. \qquad (25)$$

Here we introduced the angle-averaged moments of the complex directivity

$$\overline{R^2} = \frac{1}{4\pi} \int_{-\pi}^{\pi} d\alpha \int_{-\pi/2}^{\pi/2} d\beta \cos\beta \left| s_R(\alpha,\beta) \right|^2,$$

$$\overline{L^2} = \frac{1}{4\pi} \int_{-\pi}^{\pi} d\alpha \int_{-\pi/2}^{\pi/2} d\beta \cos\beta \left| s_L(\alpha,\beta) \right|^2, \qquad (26)$$

$$\overline{RL} = \frac{1}{4\pi} \int_{-\pi}^{\pi} d\alpha \int_{-\pi/2}^{\pi/2} d\beta \cos\beta \, s_R(\alpha,\beta) s_L(\alpha,\beta).$$

Covariance matrix **U** carries information about the intensities of the circular components of this partially coherent field, and is just another form of the covariance matrix **W**. Eq. (25), however, presents the main finding of this work. While the unknown phase $\theta$ prevent the measurement of the complex covariance of the right and left circular components, the magnitude of this correlation is measurable and carries information about the field statistics that is not included in the covariance matrix **U**. The presence of the non-zero complementary covariance matrix implies that the random complex amplitudes of the circular polarization components are not circular in general.

We introduce in-phase and quadrature parts of these complex amplitudes

$$R = R_R + iR_I, \quad L = L_R + iL_I, \qquad (27)$$

and covariance matrix of the fluctuating components of the circular amplitudes

$$\mathbf{D} = \begin{pmatrix} \langle \tilde{R}_R^2 \rangle & \langle \tilde{R}_R \tilde{R}_I \rangle & \langle \tilde{R}_R \tilde{L}_R \rangle & \langle \tilde{R}_R \tilde{L}_I \rangle \\ \langle \tilde{R}_R \tilde{R}_I \rangle & \langle \tilde{R}_I^2 \rangle & \langle \tilde{R}_I \tilde{L}_R \rangle & \langle \tilde{R}_I \tilde{L}_I \rangle \\ \langle \tilde{R}_R \tilde{L}_R \rangle & \langle \tilde{R}_I \tilde{L}_R \rangle & \langle \tilde{L}_R^2 \rangle & \langle \tilde{L}_R \tilde{L}_I \rangle \\ \langle \tilde{R}_R \tilde{L}_I \rangle & \langle \tilde{R}_I \tilde{L}_I \rangle & \langle \tilde{L}_R \tilde{L}_I \rangle & \langle \tilde{L}_I^2 \rangle \end{pmatrix}. \qquad (28)$$

Using Eqs. (24, 25) covariance matrix **D** is calculated as

$$\mathbf{D} = \frac{1}{2} \begin{pmatrix} \overline{R^2} & 0 & \mathrm{Re}\left(\overline{RL}e^{i\theta}\right) & \mathrm{Im}\left(\overline{RL}e^{i\theta}\right) \\ 0 & \overline{R^2} & \mathrm{Im}\left(\overline{RL}e^{i\theta}\right) & -\mathrm{Re}\left(\overline{RL}e^{i\theta}\right) \\ \mathrm{Re}\left(\overline{RL}e^{i\theta}\right) & \mathrm{Im}\left(\overline{RL}e^{i\theta}\right) & \overline{L^2} & 0 \\ \mathrm{Im}\left(\overline{RL}e^{i\theta}\right) & -\mathrm{Re}\left(\overline{RL}e^{i\theta}\right) & 0 & \overline{L^2} \end{pmatrix}. \qquad (29)$$

Returning to the Cartesian basis, covariance matrix **C**, Eq. (5), contains the same information as the covariance matrix **D**, and can be recovered using Eq. (17) as follows

$$\mathbf{C} = \frac{1}{4} \begin{pmatrix} \overline{R^2} + \overline{L^2} + 2\operatorname{Re}[\overline{RLe^{i\theta}}] & 2\operatorname{Im}[\overline{RLe^{i\theta}}] & 0 & -\overline{R^2} + \overline{L^2} \\ 2\operatorname{Im}[\overline{RLe^{i\theta}}] & \overline{R^2} + \overline{L^2} - 2\operatorname{Re}[\overline{RLe^{i\theta}}] & \overline{R^2} - \overline{L^2} & 0 \\ 0 & \overline{R^2} - \overline{L^2} & \overline{R^2} + \overline{L^2} + 2\operatorname{Re}[\overline{RLe^{i\theta}}] & 2\operatorname{Im}[\overline{RLe^{i\theta}}] \\ -\overline{R^2} + \overline{L^2} & 0 & 2\operatorname{Im}[\overline{RLe^{i\theta}}] & \overline{R^2} + \overline{L^2} - 2\operatorname{Re}[\overline{RLe^{i\theta}}] \end{pmatrix} \quad (30)$$

Structure of the matrix **D** indicates that each complex random amplitude $R$ and $L$ is proper, but the complex random vector $(R, L)$ is cross improper [. Meanwhile matrix **C** indicates that each component of the Jones vector $(\tilde{u}, \tilde{v})$ is improper.

Covariance matrix of the Jones vector **W**, Eq. (7) is

$$\mathbf{W} = \frac{1}{2} \begin{pmatrix} \overline{R^2} + \overline{L^2} & i\overline{R^2} - i\overline{L^2} \\ -i\overline{R^2} + i\overline{L^2} & \overline{R^2} + \overline{L^2} \end{pmatrix}, \quad (31)$$

and corresponding Stokes vector, Eq. (6) is

$$\mathbf{S} = \begin{pmatrix} \overline{R^2} + \overline{L^2} \\ 0 \\ 0 \\ -\overline{R^2} + \overline{L^2} \end{pmatrix}. \quad (32)$$

Based on the Stokes vector alone, this partially polarized wave consists of independent right and left polarized components with possibly different intensities. However, the complementary covariance matrix, Eq. (4) is

$$\mathbf{W}_C = \begin{pmatrix} \overline{RLe^{i2\theta}} & 0 \\ 0 & \overline{RLe^{i2\theta}} \end{pmatrix}. \quad (33)$$

Similar to the circular basis, covariance matrix or Stokes vector carry information about the intensities of the circular components. Non-zero complementary covariance matrix implies that the circular components are correlated, and the magnitude of their covariance can be retrieved from the complementary covariance matrix. This also suggests that the two channels of the coherent receiver can use either circular or linear polarizers.

### 4. Examples

Here we illustrate the somewhat convoluted derivation of the previous section by three simple examples of the radiation sources.

#### 4.1. Z-dipole source

Radiation zone electrical field of dipole with momentum along the $z$ – axis, [9, Sect. 9.2], can be written as

$$\mathbf{D}_Z(\mathbf{r}) = \mathrm{Re}\left[\frac{\sqrt{3}}{r^3}\begin{pmatrix} -xz \\ -yz \\ x^2+y^2 \end{pmatrix}\exp(ikr - i\omega t)\right]. \quad (34)$$

After transition to the spherical coordinates Eq. (11) takes the form

$$\mathbf{S}_Z(\alpha,\beta) = \sqrt{3}\cos\beta\,\hat{\boldsymbol{\beta}}, \quad (35)$$

Fig. 1 shows the amplitude of the monopole vector field, Eq. (34) at the nominal sphere. Field is linear polarized, with only meridional component and oscillates in phase at all directions.

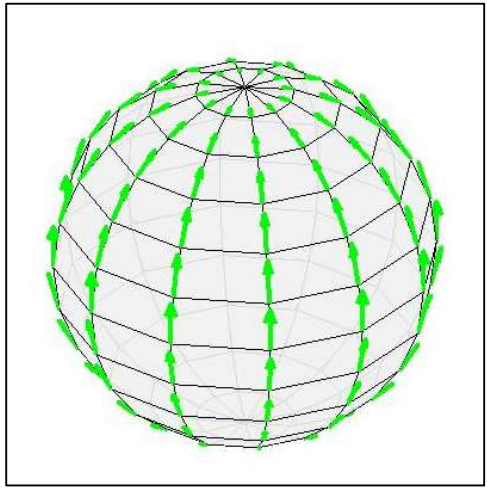

Fig. 1. Electrical field of $Z$ – dipole, Eq. (34).

Measured complex amplitudes of the circular components after transformations described by Eqs. (17 - 22) are

$$R(\alpha,\beta,\gamma) = \sqrt{\frac{3}{2}}\cos\beta\exp\left(i\theta - i\gamma + i\frac{\pi}{2}\right),$$
$$L(\alpha,\beta,\gamma) = \sqrt{\frac{3}{2}}\cos\beta\exp\left(i\theta + i\gamma - i\frac{\pi}{2}\right), \quad (36)$$

and the non-zero second moments, Eq. (26) are

$$R(\alpha,\beta,\gamma) = \sqrt{\frac{15}{2}}\sin\beta\cos\beta\exp\left(i\theta - i\gamma + i\frac{\pi}{2}\right),$$
$$L(\alpha,\beta,\gamma) = \sqrt{\frac{15}{2}}\sin\beta\cos\beta\exp\left(i\theta + i\gamma - i\frac{\pi}{2}\right), \quad (37)$$

Covariance matrix $\mathbf{W}$, and complementary covariance matrix $\mathbf{W}_C$ are

$$\mathbf{W} = \begin{pmatrix} 1 & 0 \\ 0 & 1 \end{pmatrix},\quad \mathbf{W}_C = \begin{pmatrix} 0 & e^{2i\theta} \\ e^{2i\theta} & 0 \end{pmatrix}. \quad (38)$$

Covariance matrix $\mathbf{W}$ is just a unit matrix and corresponds to an unpolarized field. However, the complementary covariance matrix $\mathbf{W}_C$ reveals the presence of "hidden" correlation between the components of the Jones vector. This correlation is evident from the Eq. (36), which shows that the random rotation along the look direction $\psi$ causes equal, but opposite sign phase changes to the circular components of the field.

### 4.2. Longitudinal ZZ-quadrupole source

Longitudinal $ZZ$ – quadrupole is just a pair of Z dipoles of the opposite polarity separated by a small distance $\Delta z$ along the $z$ axis. Radiation zone electrical field of longitudinal $ZZ$ – quadrupole, can be written in Cartesian source basis as

$$\mathbf{Q}_{ZZ}(\mathbf{r}) = \mathrm{Re}\left[\frac{\sqrt{15}}{r^4}\begin{pmatrix} xz^2 \\ yz \\ -(x^2+y^2)z \end{pmatrix}\exp(ikr - i\omega t)\right]. \tag{38}$$

After transition to the spherical coordinates Eq. (11) takes the form

$$\mathbf{S}_{ZZ}(\alpha,\beta) = \sqrt{15}\sin\beta\cos\beta\,\hat{\boldsymbol{\beta}}, \tag{39}$$

Fig. 2 shows the $ZZ$ - quadrupole vector field, Eq. (39) at the nominal sphere. Similar to the dipole case, field has only meridional component and oscillates in phase at all directions.

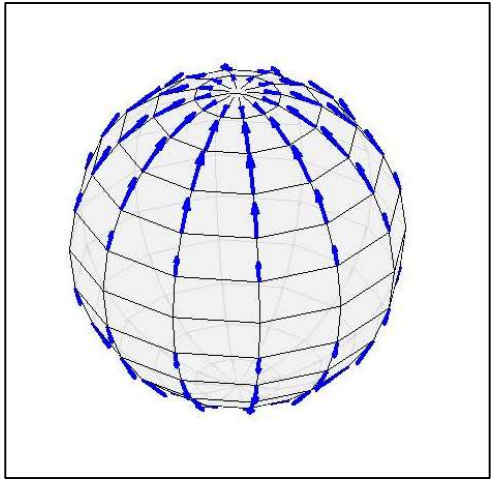

Fig. 2. Electrical field of the longitudinal quadruple, Eq. (38).

Measured complex amplitudes of the circular components after transformations described by Eqs. (17 - 22) are

$$\begin{aligned}R(\alpha,\beta,\gamma) &= \sqrt{\frac{15}{2}}\sin\beta\cos\beta\exp\left(i\theta - i\gamma + i\frac{\pi}{2}\right),\\ L(\alpha,\beta,\gamma) &= \sqrt{\frac{15}{2}}\sin\beta\cos\beta\exp\left(i\theta + i\gamma - i\frac{\pi}{2}\right),\end{aligned} \tag{40}$$

and the non-zero second moments, Eq. (26) are

$$\overline{R^2} = \overline{L^2} = \overline{RL} = 1. \tag{41}$$

Covariance matrix $\mathbf{W}$, and complementary covariance matrix $\mathbf{W_C}$ are the same as for the dipole case, Eq. (38). The rotating $\mathbf{D}_Z$ and $\mathbf{Q}_{ZZ}$ sources look identical based on the second moments of their fields even with the coherent receivers.

### 4.3. Transverse XY-quadrupole source

In order to preserve *z* as the source axis we construct the transverse *XY* quadrupole as a pair of *X* dipoles of the opposite polarity separated by a small distance $\Delta y$ along the *y* axis. Radiation zone electrical field of this quadrupole, can be written in Cartesian source basis as

$$\mathbf{Q}_{XY}(\mathbf{r}) = \mathrm{Re}\left[\sqrt{\frac{24}{5}}\frac{1}{r^4}\begin{pmatrix} y(y^2+z^2) \\ -xy^2 \\ -xyz \end{pmatrix}\exp(ikr - i\omega t)\right]. \tag{42}$$

After transition to the spherical coordinates Eq. (11) takes the form

$$\mathbf{S}(\alpha,\beta) = \sqrt{\frac{24}{5}}\cos\beta\sin\beta\,\hat{\boldsymbol{\beta}} - \sin\alpha\,\boldsymbol{\alpha}. \tag{43}$$

Fig. 3 shows the *XY* - quadrupole vector field, Eq. (39) at the nominal sphere. Field is still linear polarized and in phase for all directions. Unlike the dipole and longitudinal quadrupole cases, polarization direction varies with the azimuthal angle.

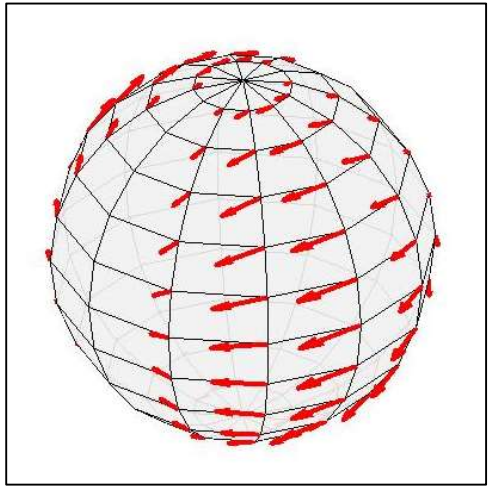

Fig. 3. Electrical field of the lateral quadruple, Eq. (42.).

Measured complex amplitudes of the circular components after transformations described by Eqs. (17 - 22) are

$$\begin{aligned}R(\alpha,\beta,\gamma) &= \sqrt{\frac{12}{5}}\left(-\sin^2\alpha\cos\beta + i\cos\alpha\sin\beta\right)\exp(i\theta - i\gamma), \\ L(\alpha,\beta,\gamma) &= \sqrt{\frac{12}{5}}\left(-\sin^2\alpha\cos\beta - i\cos\alpha\sin\beta\right)\exp(i\theta - i\gamma),\end{aligned} \tag{44}$$

and the non-zero second moments, Eq. (26) are

$$\overline{R^2} = \overline{L^2} = 1,\ \overline{RL} = \frac{1}{5}. \tag{45}$$

Covariance **W** and complementary matrices are

$$\mathbf{W} = \begin{pmatrix} 1 & 0 \\ 0 & 1 \end{pmatrix},\quad \mathbf{W}_C = \frac{1}{5}\begin{pmatrix} 0 & e^{2i\theta} \\ e^{2i\theta} & 0 \end{pmatrix}. \tag{46}$$

As was stated above, only the magnitude $\overline{|RL|}$ is measurable, and in this case, it indicates the more complex angular distribution of the radiated field of the lateral quadrupole.

5.  **Summary**

Based on the complete second-order statistics of the narrowband electromagnetic field [5] we demonstrated that the field created by a randomly rotated source is an improper complex random vector. As a result, the statistics of this field is not completely described by the common Stokes vector or covariance matrix.

The additional statistical parameter of this improper field is related to the complementary covariance matrix. It can be measured by a two-channel coherent receiver and potentially it carries additional information about the source.

We examined three simple source examples and calculated the elements of both covariance matrices. The standard covariance matrices are not able the recognize the differences between the sources. However, by invoking the complementary covariance matrix it is possible to recognize the difference between the lateral quadrupole and the dipole or longitudinal quadrupole.

The natural extension of this work would be investigation of the two-points coherence matrix and associated two-points complementary covariance matrix.

It would be also more practical to examine the complete statistics of the field scattered by a rotating object with application to coherent radars and remote sensing.